\documentclass[sigconf]{acmart}
\usepackage{tabularx}
\usepackage{algorithm}
\usepackage{algorithmic}
\usepackage{etoolbox}
\usepackage{xcolor}
\newtoggle{showadditions}
\newtoggle{showdeletions}
\toggletrue{showadditions}
\togglefalse{showdeletions}

\AtBeginDocument{%
  }

\copyrightyear{2025}
\acmYear{2025}
\setcopyright{cc}
\setcctype{by}
\setcctype{by}
\acmConference[ASSETS '25]{The 27th International ACM SIGACCESS Conference
on Computers and Accessibility}{October 26--29, 2025}{Denver, CO, USA}
\acmBooktitle{The 27th International ACM SIGACCESS Conference on Computers
and Accessibility (ASSETS '25), October 26--29, 2025, Denver, CO, USA}
\acmDOI{10.1145/3663547.3746387}
\acmISBN{979-8-4007-0676-9/2025/10}

\begin{document}

\title[ADHD Content Characteristics \& Quality on Video Sharing Platforms]{Characterizing Collective Efforts in Content Sharing and Quality Control for ADHD-relevant Content on Video-sharing Platforms}

\author{Hanxiu `Hazel' Zhu}
\affiliation{%
  \institution{University of Wisconsin-Madison}
  \city{Madison}
  \state{Wisconsin}
  \country{USA}
}
\email{hzhu339@wisc.edu}

\author{Avanthika Senthil Kumar}
\affiliation{%
  \institution{University of Wisconsin-Madison}
  \city{Madison}
  \state{Wisconsin}
  \country{USA}
}
\email{asenthilkum5@wisc.edu}

\author{Sihang Zhao}
\affiliation{%
  \institution{The Chinese University of Hong Kong, Shenzhen}
  \city{Shenzhen}
  \country{China}
}
\email{222010544@link.CUHK.edu.cn}

\author{Ru Wang}
\affiliation{%
  \institution{University of Wisconsin-Madison}
  \city{Madison}
  \state{Wisconsin}
  \country{USA}
}
\email{ru.wang@wisc.edu}

\author{Xin Tong}
\affiliation{%
  \institution{ The Hong Kong University of Science and Technology (Guangzhou)}
  \city{Guangzhou}
  \country{China}
}
\email{xint@hkust-gz.edu.cn}

\author{Yuhang Zhao}
\affiliation{%
  \institution{University of Wisconsin-Madison}
  \city{Madison}
  \state{Wisconsin}
  \country{USA}
  }
\email{yuhang.zhao@cs.wisc.edu}

\renewcommand{\shortauthors}{Zhu et al.}

\begin{abstract}
  Video-sharing platforms (VSPs) have become increasingly important for individuals with ADHD to recognize symptoms, acquire knowledge, and receive support. While videos offer rich information and high engagement, they also present unique challenges, such as information quality and accessibility issues to users with ADHD. However, little work has thoroughly examined the video content quality and accessibility issues, the impact, and the control strategies in the ADHD community. We fill this gap by systematically collecting 373 ADHD-relevant videos with comments from YouTube and TikTok and analyzing the data with a mixed method. Our study identified the characteristics of ADHD-relevant videos on VSPs (e.g., creator types, video presentation forms, quality issues) and revealed the collective efforts of creators and viewers in video quality control, such as authority building, collective quality checking, and accessibility improvement. We further derive actionable design implications for VSPs to offer more reliable and ADHD-friendly content. 
\end{abstract}


\begin{CCSXML}
<ccs2012>
   <concept>
       <concept_id>10003120.10003130.10011762</concept_id>
       <concept_desc>Human-centered computing~Empirical studies in collaborative and social computing</concept_desc>
       <concept_significance>500</concept_significance>
       </concept>
   <concept>
       <concept_id>10003120.10011738.10011773</concept_id>
       <concept_desc>Human-centered computing~Empirical studies in accessibility</concept_desc>
       <concept_significance>500</concept_significance>
       </concept>
 </ccs2012>
\end{CCSXML}

\ccsdesc[500]{Human-centered computing~Empirical studies in accessibility}
\ccsdesc[500]{Human-centered computing~Empirical studies in collaborative and social computing}

\keywords{video-sharing platforms, information quality, online communities, ADHD, accessibility}
\maketitle

\section{Introduction}
Attention Deficit Hyperactivity Disorder (ADHD) has received increasing attention in recent years as more individuals recognize their symptoms and seek diagnosis and support. Despite the growing awareness, misconceptions about ADHD persist, such as the belief that it affects only White boys \cite{Bruchmller2012} or that individuals have to exhibit hyperactivity \cite{LAUFER1957463}. %Exacerbated by racial and gender biases in clinical trial representation, ADHD researchers and clinicians tended to pay more attention to White boys \cite{Bruchmller2012}, 
%overlooking the diverse manifestations of ADHD in adults, females, and people of color, who may exhibit less noticeable but highly prevalent inattentive symptoms \cite{Slobodin2020}. 
These misconceptions leave many people unaware of their ADHD, and even people who recognize their symptoms often face difficulties in obtaining proper diagnosis and treatment \cite{Attoe2023}. As a result, information and resources online have been crucial for individuals with ADHD to identify and understand their symptoms and experiences and seek support from peers \cite{eagle2023}.

% various social media paltforms are used by people with adhd; different types of social media, among which video-based could be particularly useful

% why we are focusing on video-based platforms; benefits and drawbacks of videos; different forms of videos (longer and shorter); we aim for higher coverage and look comprehensively look into different types of platforms; people with adhd use both tiktok and youtube

Among various social media platforms, \textit{video-sharing platforms (VSPs)}, such as TikTok and YouTube, have become an emerging medium for people with ADHD to share experiences and exchange information \cite{Thapa_Thapa_Khadka_Bhattarai_Jha_Khanal_Basnet_2018}.
%Discussions of ADHD on social media have increased tremendously in the recent few years, especially since the sharp growth of social media use during the COVID-19 lockdown \cite{Werling2021, Xu2023-vs}. People with ADHD have been using a wide range of social media platforms to exchange health information, from the conventional text-based forums \cite{Fleischmann2012} to platforms that enable multi-modal sharing of text, images, and videos \cite{Kang2016, Gajaria2011}. 
Unlike conventional text- or image-based media, videos offer richer information via multimodal channels, enabling increased engagement, higher persuasiveness, and more effective behavior intervention \cite{Li2022, Wittenberg2021}. This has led to a surge in the ADHD audience as well as ADHD-relevant content on VSPs. For example, by June 2024, TikTok had more than three million video posts and 25 billion views under the hashtag ``\#adhd,'' highlighting the influence of VSPs on ADHD-relevant discussions. 

Recent research has started examining %how online ADHD community members create, evaluate, and respond to 
the experiences of ADHD users with VSPs \cite{Leveille2024, eagle2023, McDermott2022}. For example, \citeauthor{eagle2023} \cite{eagle2023} conducted a digital ethnography study to analyze ADHD-relevant posts and comments on Twitter, Instagram, and TikTok, and identified TikTok as a valuable source of shared expertise for people with ADHD. These works mainly focused on the community-building and content-sharing experiences of ADHD users, without deeply investigating the potential content quality and delivery issues brought by the unique video form. 

Video content can bring unique risks and challenges to users with ADHD. The high information richness and less controllable content flow in videos make it particularly challenging to identify and combat misinformation \cite{Niu2023}, potentially leading to hasty ADHD self-diagnosis and misunderstanding \cite{Gilmore_Beezhold_Selwyn_Howard_Bartolome_Henderson_2022}. % \yuhang{a couple of sentences with references to reveal risks, e.g., information quality is harder to discern and control, etc.} 
The multimodal nature of videos could also bring unique perceptual challenges to ADHD viewers \cite{Nikkelen2014MediaUA}.
To reveal the content quality issues on VSPs, \citeauthor{Yeung_Ng_Abi-Jaoude_2022} \cite{Yeung_Ng_Abi-Jaoude_2022} and \citeauthor{Thapa_Thapa_Khadka_Bhattarai_Jha_Khanal_Basnet_2018} \cite{Thapa_Thapa_Khadka_Bhattarai_Jha_Khanal_Basnet_2018} examined the prevalence of misleading ADHD content on TikTok and YouTube respectively by asking medical experts to rate video quality using existing or self-devised information quality frameworks. %\yuhang{describe the work directly. did they ask experts to label the videos?} 
However, both works focused on quantitative measures through a medical lens, overlooking the complexity and nuances of user reactions, challenges, and strategies in video quality control. % which can further uncover user experiences and inspire future technology design to monitor and enhance video content quality and accessibility on VSPs. 
With the increasing amount of ADHD-relevant videos online, it is critical to deeply investigate the characteristics of such videos on VSPs, their content quality and accessibility issues for ADHD users, and how ADHD creators and viewers respond to and combat these issues, thus inspiring a more inclusive, safe, and trustworthy VSP community for ADHD. % is delivered using videos and how people are responding to it. 

%\added{While their work involved exploring ADHD community building on TikTok, they didn't focus on videos as a unique form of presentation and the underlying challenges related to video qualities.}
%\yuhang{what did they find about TikTok?} 
%While involving TikTok as one study platform, they investigated all types of social media as a whole and did not focus on the unique characteristics and impact of shared videos on users with ADHD. % without fully exploring how people convey and evaluate information with different forms of videos. 
%Eagle et al. conducted a digital ethnography study on ADHD-relevant posts and comments on Twitter, Instagram, and TikTok \cite{eagle2023}, qualitatively analyzing people's experiences through online communities and identifying TikTok as a valuable source of shared expertise for people with ADHD. 
%\yuhang{what did they find about TikTok?} 
%While involving TikTok as one study platform, they investigated all types of social media as a whole and did not focus on the unique characteristics and impact of shared videos on users with ADHD. % without fully exploring how people convey and evaluate information with different forms of videos. 
%There is a gap in understanding how various stakeholders---creators, viewers, and the platforms---engage in the video content generating, sharing, and viewing on the emerging video sharing platforms. % about the presentation, perspective, and quality of the ADHD-related videos that they create and consume.
Our research focuses on emerging VSPs and investigates the unique characteristics and quality issues of ADHD-relevant videos. We contextualize our research on the two most popular and representative VSPs---\textit{TikTok} that shares short video content and \textit{YouTube} that mainly offers long video content. %\footnote{
%\added{YouTube introduced Shorts in 2021. However, it is relatively new and is not as widely viewed as TikTok \cite{Thang2023}. Thus, we focused on examining standard-length videos on YouTube to derive more distinctive and meaningful comparison.}}. 
We address the problems from both the content creators' (via the videos) and viewers' (via the comments) perspectives, unfolding their challenges, strategies, and social dynamics in monitoring, controlling, and improving video quality and accessibility on VSPs. 
%share experiences, disseminate resources, identify issues, and develop communities. 
Specifically, we answer the research questions below:   
% examination of creator information, videos and comments on TikTok and YouTube, yielding both quantitative and qualitative findings. Unlike other social media platforms that also afford text and images for creators to deliver their ideas, TikTok and YouTube focus on presenting videos of different lengths, with text as supplements.  Given this specific format, what have creators and viewers done in conveying ADHD-related information, and how could video-based platforms better facilitate such information exchange? Towards this end, we answer the following questions: 

\begin{enumerate}
\item What types of ADHD-relevant content are covered on YouTube and TikTok respectively? What are their characteristics (e.g., creator types, presentation forms) and how do they impact content quality? 

\item How do content creators establish authority and trust on their videos and maintain video quality? % \added{in terms of contents and delivery}?

\item How do viewers of ADHD-relevant videos perceive, interpret, and respond to different video creators, content, and qualities?
\end{enumerate}

To answer the questions, we systematically collected and analyzed %8946 YouTube and 50085 TikTok ADHD-relevant videos, and sampled 
373 videos with comments (YouTube: 189, TikTok: 184) for an in-depth content analysis. We analyzed the data using a mixed method to uncover both quantitative characteristics and distributions of ADHD-relevant videos and qualitative understanding of user challenges and strategies. We found that TikTok and YouTube have distinctive creator demographics and presentation forms (e.g., more personal ADHD experience sharing by individual creators on TikTok vs. more educational and medical knowledge by health practitioners and organizations on YouTube), serving different content sharing purposes yet posing different issues.  Through examining the collective quality control efforts, we highlighted potential challenges in these quality control practices (e.g., risks of misleading authority) and their impact on viewers, uncovering unique creator-viewer dynamics (e.g., creators reminding viewers to seek clinical help vs. viewers complaining the lack of accessible diagnostic resources for adult women with ADHD) in ADHD communities on VSPs. Furthermore, we discovered video accessibility issues that largely affect ADHD viewers' ability to intake the video content, including video length, slow pace, distracting multimodal elements, and caption availability, and revealed the corresponding practices and challenges when combating these issues. %ADHD accessibility practices on VSPs, revealing their promises and challenges.
    
Our contributions are twofold. First, to the best of our knowledge, this is the first research that conducts both qualitative and quantitative examination of ADHD-relevant videos on VSPs with a focus on content quality and accessibility. Our research focuses on the collective efforts of the creators, viewers, and the platforms, revealing current challenges and practices around trust building and quality control. While our study focused on ADHD, the insights can potentially be generalized to the broader neurodivergent or mental health communities. Second, based on the characterization and analysis of ADHD-relevant content and user interactions, we derived concrete and actionable design implications to enable more inclusive video-watching experiences and facilitate quality control and accessibility efforts for the ADHD community.
    
    %(1) more effective content quality control, and (2) more accessible ADHD video creation, inspiring more trustworthy and inclusive VSP communities. 
    % \yuhang{need adjustment as your discussion is not structured in this way anymore} 

\section{Background \& Related Work}
% both videos and comments
% talk about how other disorders have been examined on other social media platforms
% explicitly mention the gap

In this section, we explain the historically ingrained diagnostic challenges and the importance of active help-seeking for individuals with ADHD. We also examine the role of video-sharing platforms in supporting help-seeking and community-building within the ADHD community, as well as the underexplored issue of ADHD-relevant video qualities on these platforms.

% write the introduction first
\subsection{ADHD: Diagnostic Challenges}
\label{sub-sec:related-work:adhd}
% not highly relevant; trim
% diagnosis paragraph is important; relate it to how people leverage social media for health information
% be very explicit about how readers should interpret work; draw dots; don't leave it for readers to interpret
% don't criticize work who did not focus on people with adhd

Attention-deficit/hyperactivity disorder (ADHD) is a neurodevelopmental disorder that affects 7.6\% of children and 6.8\% of adults \cite{Salari2023}. People with ADHD could display inattention and/or hyperactivity/impulsivity \cite{Wilens2010-hq}, and poorly managed ADHD could put people under various risks, including low self-esteem, underachievement, and substance abuse \cite{Watters2017}. 

Despite the prevalence of ADHD and the importance of timely recognition and treatment, diagnosing ADHD can be challenging. Previously known as hyperkinetic behavior syndrome, ADHD was originally characterized as a childhood disorder that mainly affects boys who exhibit disruptive behaviors \cite{LAUFER1957463}. %Such a characterization, and thus the stereotypes associated, overlooked the inattentive sub-type of ADHD, preventing 
As a result, children who have the inattentive sub-type of ADHD have more difficulties getting a timely diagnosis and intervention in childhood \cite{simon_czobor_bálint_mészáros_bitter_2009}. 
%Moreover, the high rate of comorbidity of ADHD with other disorders could further complicate the diagnosis \cite{Spencer2007, Koyuncu2022}. Additionally,
Additionally, biases along the axes of age, gender, and race also contribute to diagnostic challenges. Adults \cite{Ginsberg2014, Newcorn2007, RivasVazquez2023}, women \cite{Young2020, Kelly2024}, people of color \cite{Shi2021, Zhao2023}, and their intersections \cite{Morgan2023, Attoe2023, Waite2009} are more likely to experience undiagnosed or misdiagnosed ADHD, as their experiences are misinterpreted or dismissed \cite{Attoe2023, Holthe2017-ah}. For individuals who failed to have their ADHD diagnosed early, they gradually adopt coping strategies or masking techniques to fit into their environment, making their ADHD presentations even less identifiable \cite{Kosaka2018, AgnewBlais2016}. 

As a result, self-exploration and awareness become crucial for people with undiagnosed ADHD to realize the need to seek help \cite{Pawaskar2019, Lee2023}. Social media thus becomes an indispensable source for users with ADHD. While ADHD affects each individual differently, research has shown that people with ADHD could be particularly attracted to social media \cite{Andreassen2016}, a condition that was exacerbated during the COVID-19 lockdown \cite{Werling2021}. We unpack the impact of social media on ADHD and the broader health community below.

%The HCI community has endeavored to understand the living experiences of people with ADHD and other neurodivergent members \cite{remotework, Desrochers}, and design tools that help people and families with ADHD navigate their daily lives \cite{10.1145/2858036.2858157, 10.1145/3313831.3376837}. Some work has also tapped on facilitating diagnosis \cite{Bautista_Hernandez-Vela_Escalera_Igual_Pujol_Moya_Violant_Anguera_2016, temporalhashing} and intervention \cite{10.1145/2661334.2661399, 10.1145/2811681.2824997}, with the aim of identifying or ``alleviating'' ADHD symptoms. Nonetheless, many work proposing designs or technologies had little involvement of people with ADHD, but instead consulted parents, teachers, healthcare providers and other ADHD ``experts'' for inspirations and feedback \cite{Spiel}. While this criticism stems from the perspective of user-centered design, it is important to recognize and magnify the voices of individuals with ADHD in other contexts of HCI research. 

\subsection{Using Social Media for Health Purposes}

% highlight most important related work; little information about how people with ADHD leverage social media except recent work; highlight what they did and identify what they did not do

% highlight that videos only boom recently; no prior work; and they are advancing so quickly
% Social media serves as an important place for users with health needs \cite{pretoriushelpseeking, Choudhury_De_2014}. For example, \textit{Facebook} allows users to form both public \cite{Afsar2024} and private groups \cite{Kelly2022} to create a common space while protecting users' privacy. \textit{Reddit} allows users to create sub-reddit communities where users interact anonymously, encouraging more candid sharing on sensitive topics, such as suicide ideation and support \cite{Henninger2019, SilveiraFraga2018, RedondoSama2021}. \textit{Twitter} and \textit{Instagram} utilizes the hashtag feature as means of forming unbounded community, enabling rapid information spread and wide coverage of health education and advocacy \cite{Dzienny2022, Santarossa2018}.
Conventional text/image-based social media platforms, including Facebook, Reddit, and Twitter, have long been used to support users with health needs in sharing experiences, exchanging support, and deconstructing stereotypes \cite{Gajaria2011, Thelwall2020, Schaadhardt2023}. %\yuhang{what about research for ADHD on non-video social media?}
More recently, however, \textbf{video-sharing platforms (VSPs)} have emerged rapidly among health communities, since videos afford particularly rich information and fast propagation \cite{Li2014video}. \textit{YouTube} and \textit{TikTok} are the two most prominent VSPs in the U.S., with 83\% of adults reported using YouTube and 33\% using TikTok \cite{pew2024adults}. Both platforms bring increasing impact on the dissemination of a wide range of health-related content \cite{Madathil2014}. 

Compared to other media forms, videos can more easily garner neurodivergent users' attention and provide a greater sense of enjoyment \cite{Nikkelen2014MediaUA}. %\yuhang{add something like videos are particularly preferred by ADHD users since xxx.}. 
Recently, there has been a surge in ADHD-relevant content on VSPs, particularly following the significant growth in short-form video viewers during COVID-19 \cite{Xu2023-vs}. Creators have published more than three million posts under the hashtag ``\#adhd'' on TikTok with more than 25 billion views (as of June 2024), making it one of the most popular health-related hashtags \cite{Zenone2021}. 

Despite the increasing popularity of VSPs and the surge of ADHD content, the characteristics of videos and their impact on the ADHD community remain underexplored. Limited prior work has started investigating how people with ADHD use VSPs to construct ADHD discourse and form communities \cite{Kang2016, McDermott2022}. For example, \citeauthor{Leveille2024} \cite{Leveille2024} analyzed the personal and humorous nature of ADHD-relevant content on TikTok, unveiling their role in reflecting neurodivergent identity. \citeauthor{eagle2023} \cite{eagle2023} conducted a digital ethnography to examine ADHD communities across TikTok, Twitter, and Instagram, highlighting VSPs as valuable sources of information and support. Moreover, prior work has also started to explore neurodiverse users' experiences with the video watching and creation process \cite{mcdonnell2024}. For example, \citeauthor{adhdvideoaccess} \cite{adhdvideoaccess} conducted an interview study to unveil the challenges and strategies of viewers with ADHD when consuming video contents. \citeauthor{simpson2023} \cite{simpson2023} explored the experiences of neurodiverse TikTok creators with an interview study, revealing the need for VSPs to support such creators. However, while prior work has contributed to qualitative understanding of ADHD community building, advocacy and creative experiences on VSPs, none of these works has examined the different \textit{characteristics} and \textit{qualities} of ADHD-relevant videos and their impacts on viewers. In contrast, our work examines the ADHD communities on different VSPs from the perspective of content quality and accessibility, which are critical towards building a safer and more inclusive online environment for users with ADHD.

%\added{However, even though such work has involved qualitatively examining content from a \textit{single} VSP, they focused on understanding how ADHD community members constructed the ADHD discourse and formed support networks on VSPs. Our work, however, focused on another important aspect of VSPs for the ADHD communities---video qualities---and sought to understand the videos, their presentation forms and users' interaction dynamics in-deph. } \yuhang{rationale is still not strong...prior work has focused on either YouTube or TikTok for ADHD, then what's new for us?}.
%\yuhang{this section should be talking about prior research on ADHD+video sharing platforms, e.g., Eagle et al.}
% previous work has started investigating how adhd build community, including [...]. For example, [...]. However, they didn't focus on [...]. 

\subsection{Quality of Health Contents on Social Media} \label{informationAssessment}
% consider merge this with 2.2
% quality benchmarks might not be necessary (need to highlight what are the benchmarks lacking); put under method instead of related work
% highlight the gap
% unclear what's the most relevant work!! need to highlight the most relevant work, unpack their methodology and findings, point out what they are lacking
% who did that; how they did that; what's their key findings; what they are lacking and how we contribute to the gap

% problem of health misinformation on social media

Despite the important role that social media plays in offering health-relevant content, research has called the quality of such content into question. Numerous efforts in the health field have been made to assess the quality of health information on various social media platforms, covering topics of drugs \cite{Allem2017, Yang2018}, vaccines \cite{Basch2017, Schmidt2018}, chronic illnesses \cite{Biggs2013, Mueller2019}, pandemics \cite{Ahmed2019, Gabarron2021}, and mental health \cite{Starvaggi2024, LorenzoLuaces2023}, and discovered that 30\% to 87\% of posts on different social media platforms contain misinformation \cite{SuarezLledo2021}. %Researchers in health field commonly define misinformation as false, inaccurate, incomplete, misleading, or non-scientific-evidence-supported information \cite{SuarezLledo2021, Chou2018, Pan2021}. 
%For example, a systematic review in 2021 examined 69 papers that assessed the quality of health information on various social media platforms and discovered that 30\% to 87\% of posts on social media platforms contained misinformation \cite{SuarezLledo2021}. 

% problem of health misinformation for ADHD on VSPs
The content quality issues for ADHD-relevant videos on VSPs can be more severe and challenging due to the rich, multimodal, and less controllable nature of videos \cite{Niu2023}. For example, \citeauthor{Thapa_Thapa_Khadka_Bhattarai_Jha_Khanal_Basnet_2018} \cite{Thapa_Thapa_Khadka_Bhattarai_Jha_Khanal_Basnet_2018} collected ADHD-relevant videos on YouTube, rated the video quality with a self-devised scoring system, and identified 38.4\% of the videos as misleading. %\yuhang{how did they identify it? This is important details since we are using a different approach.}
More recently, \citeauthor{Yeung_Ng_Abi-Jaoude_2022} \cite{Yeung_Ng_Abi-Jaoude_2022} invited experts to rate the top 100 videos under the hashtag ``\#adhd'' on TikTok in 2022, and identified 52\% as misleading. 

% In addition to assessing information quality of health-relevant contents, researchers have also sought to understand how social media users manage health information of different qualities. Prior work has recognized the important roles played by different stakeholders in combating health misinformation, including signaling misinformation in comments \cite{Bode2017, Seo2022, Zhang2024}, having healthcare practitioners as creators to debunk misinformation \cite{Royan2022, Bautista2021, Neylan2021}, and having social media platforms recognize health misinformation via manual \cite{juneja2022} or automatic fact-checkers \cite{Barve2021, Barve2023}. For VSPs, research has also looked into generalizing specific elements in video (e.g., creators' tone evoking strong emotions) to detect and signal misinformation \cite{Hartwig2024}. However, none of these work has specifically focused on ADHD communities on VSPs. ADHD-relevant videos pose unique challenges to content quality control due to the highly comorbid (65\% to 89\%) and diverse ADHD conditions \cite{reale2017, Sobanski2006}, the vulnerability of experiences and emotions delivered in videos \cite{iseemehere}, and the tension between the medical and the ADHD communities \cite{eagle2023}. These challenges make ADHD-relevant video quality assessment and control a particularly subtle task that requires deeper thoughts and examination. Our work contributes to this line of research by characterizing how creators and viewers of ADHD-relevant videos evaluate and manage video qualities.
However, although prior work has highlighted the potential content quality risk of ADHD-relevant videos, such quantitative evaluation of ADHD video quality relied on a score-based assessment system developed from a clinical lens \cite{Azer2020-lg}, without diving into creators' and viewers' reactions, practices, and challenges when combating video quality issues. Such clinical criteria are usually not well-suited for videos that share unique personal experiences---an important type of content for ADHD community building and experience sharing \cite{eagle2023}. This gap highlights the need for a more nuanced and thorough approach to examine the characteristics, qualities, and impact of ADHD content on VSPs.

% Aside from examining information quality from a professional perspective, recent HCI work has also looked into how social media users assess information. Via an interview study, Milton et. al. \cite{milton2024} found that users on social media leverage various factors, including information ranking, repetition, and creators' profiles, to decide if a piece of information can be trusted. Meanwhile, some users take an additional step beyond assessing information. Eagle et al. conducted a digital ethnographic study to examine ADHD-related content on Twitter, Instagram, and TikTok, they gave examples of psychiatrists and psychology professors on TikTok creating videos and comments to challenge misleading videos \cite{eagle2023}. This finding provides valuable insights into how authoritative figures within an ADHD community could proactively safeguard other community members from problematic information. However, as Eagle et al. also pointed out in their work, the constitution of ADHD community is complicated, with members of various roles coming from diverse backgrounds. Our work seeks to contribute to a deeper understanding of how different ADHD community members on YouTube and TikTok strive to improve the information quality of their unbounded communities, the challenges they are facing, and how platforms could better facilitate their efforts.

% need to highlight challenges for video-sharing platforms, what current work has done \cite{Niu2023,  Hartwig2024}, and what is missing

\section{Methodology}
Our work seeks to understand how ADHD community members on VSPs share, consume, and evaluate content. We collected and analyzed both ADHD-relevant videos and comments to understand video characteristics and user interaction dynamics. We elaborate on our choices of platforms, data collection and analysis process below.  

\subsection{Platform Choices} We chose YouTube and TikTok, the two most popular VSPs in the United States \cite{pew2024adults}. Both platforms engaged heated discussions on ADHD and were criticized for containing content of varying quality \cite{Thapa_Thapa_Khadka_Bhattarai_Jha_Khanal_Basnet_2018, Yeung_Ng_Abi-Jaoude_2022}. While YouTube introduced \textit{Shorts} in 2021, it is relatively new and is not as widely viewed as TikTok \cite{Thang2023}. Thus, we focused on examining standard-length videos on YouTube to derive more distinctive comparison.

% The general procedure of our work is as follows: First, we scraped publicly available videos and comments from TikTok and YouTube, and sampled videos for analysis. We then conducted open-coding samples, and applied [...] to generate themes and findings. We detail each process in the following sub-sections. 

% overview of process: collection videos and comments, analyze data

\subsection{Data Collection \& Sampling} We systematically and exhaustively collected videos, their metadata, and comments from TikTok and YouTube. Additionally, we collected video descriptions and creators' profiles, which contain important resource and creator information. We conducted video search with the platform's default priority settings (i.e., relevance\footnote{\textit{YouTube}: \url{https://support.google.com/youtube/answer/111997}; \textit{TikTok}: \url{https://support.tiktok.com/en/using-tiktok/exploring-videos/discover-and-search}}) to collect videos that are more likely to be recommended to users. We specify the video collection methods for each platform below. %\yuhang{What is the general criteria and goal of the data collection? E.g., are we trying to be exhaustive? are we trying to control anything to ensure consistency between the two platforms? What other data did we collect and why? e.g., comments, meta data from the video, e.g., what hashtags, creator profile etc.}

\subsubsection{TikTok Video Collection.}
Following prior work \cite{wang2023autistic}, we collected ADHD-relevant videos on TikTok with \textit{Apify}, a third-party scraping tool\footnote{ \textit{Apify}: \url{https://apify.com/clockworks/tiktok-scraper.}}. We did not use the official TikTok API as it did not support any priority search (e.g., results ranked by relevance). To ensure comprehensive coverage of videos, we used hashtags as the search index due to their primary influence on video search and organization on TikTok \cite{Anderson2020}, following prior research \cite{Ling2023}. 

We started by searching the hashtag ``\#adhd'' and collected 3984 videos. Using the search results as seeds, we further identified other relevant hashtags. Specifically, we extracted and ranked the top 100 most frequent hashtags that co-occur with ``\#adhd'' from the 3984 videos, manually checked and removed those that are not directly related to ADHD (e.g., \#fyp, \#ocd, \#lgbtq), and generated a new set of hashtags as search terms for another round of video search. %to find other frequently co-occurring ADHD-related hashtags.
We repeated this process until the hashtag set saturated. In the end, we obtained 55 hashtags (see Table \ref{tab: datacollection}). For each hashtag, we scraped the top 1000 videos, and collected a total of 50085 videos on November 10th, 2023. After removing the duplicated videos, we had 34067 unique videos. As comment analysis is a critical part of our research to understand viewers' responses and interactions between creators and viewers, we dropped videos with fewer than 20 comments, resulting in 26318 TikTok videos.

\subsubsection{YouTube Video Collection.} \label{youtubeCollection} We used the YouTube Data API  %\footnote{\url{https://developers.google.com/youtube/v3/docs}} 
to collect videos from YouTube. While hashtagging is available on YouTube, it is not the primary way to relate and organize videos on YouTube \cite{Tong2022}. We thus employed a keyword-based search approach and strived to derive a comprehensive set of keywords to cover diverse ADHD-related topics. To identify the keywords, we referred to the TikTok dataset to inform popular ADHD-relevant topics. Specifically, we identified the top 1000 most frequently occurring hashtags from the TikTok dataset. Two researchers then discussed and grouped them into 26 categories (see Table \ref{tab: datacollection}). We used these categories as keywords for YouTube video search, combining “ADHD” with each category keyword (e.g., ADHD parenting) as the search term to retrieve videos on each topic. We collected videos from each search term until the search results became irrelevant \cite{komoaite2019}. This approach resulted in 8946 videos (on November 28th, 2023). After removing the duplicated videos, we had 5285 unique videos and then narrowed down to 2281 having more than 20 comments. 

\begin{table}[h]
%\scriptsize 
\centering
\caption{Search terms used for collecting TikTok and YouTube videos.}
\resizebox{0.98\columnwidth}{!}{
\begin{tabularx}{\columnwidth}{p{0.95\columnwidth}}
\toprule
\textbf{TikTok Hashtags}\\
\midrule 
adhd, 
adhdtok, 
adhdbrain,  
adhdsquad, 
adhdtiktok, 
adhdexplained, 
adhdcommunity, 
80hd, 
add, 
adhdanonymous, 
adhdtribe, 
actuallyadhd, 
neurodivergent, 
neurospicy, 
neurodivergenttiktok, 
adhdlife, 
adhdinwomen, 
adhdwomen, 
womenwithadhd, 
adhdgirls, 
adhdingirls, 
adhshe, 
adhdinmen, 
adultadhd, 
adhdadult, 
adhdinadults, 
adhdcheck, 
adhddiagnosis, 
adhdsymptoms, 
latediagnosisadhd, 
adhdprobs, 
adhdproblems, 
adhdstruggles, 
adhdtips, 
adhdhelp, 
adhdsupport, 
adhdhacks, 
adhdcoach, 
adhdcoaching, 
adhdtipsandtricks, 
adhdmemes, 
adhdhumor, 
adhdcouple, 
adhdrelationships, 
adhdpartner, 
adhdmom, 
adhdparenting, 
adhdfamily, 
audhd, 
audhder, 
audhdtok, 
audhdtiktok, 
adhdawareness, 
adhdisreal
\\ 
\midrule
\textbf{YouTube Categories}\\
\midrule 
parenting, relationship, workplace, school \& education, gender, race, sexuality, adult, kids, problems \& challenges, knowledge \& facts, diagnosis, symptoms, comorbidity, mental health, physical health, help \& tips, tools \& technologies, community, medication \& treatment, positivity,	creator, memes, awareness \& advocacy, stigma, life \& experiences
%\\
%\textit{* Two keywords connected by ``\&'' belong to the same category.}
\\ 
\bottomrule
\end{tabularx}}
\label{tab: datacollection}
\end{table}

\subsubsection{Video Sampling.}
While collecting a comprehensive set of videos, we also needed to further narrow down the scope to get a reasonable number of videos for analysis. We thus applied a critical case sampling approach \cite{Patton1999-dk} to achieve a balanced video sampling across all ADHD topics. However, as search results from YouTube rely on both the keywords and video popularity and quality \cite{youtube_howyoutubeworks_search}, the returned results may not always match the intended search topic (e.g., when searching \textit{ADHD parenting}, a popular ADHD video that is unrelated to parenting might be returned). To address this issue, we conducted another round of categorization to more accurately assign each video to the appropriate categories in Table \ref{tab: datacollection}. TikTok videos were categorized based on whether a video contained hashtags that belonged to a certain category. YouTube videos were manually categorized by analyzing their titles and descriptions: Two researchers independently categorized 200 randomly selected videos, discussed their categorization to resolve disagreements, and split the remaining videos to complete all video categorization. 

We selected 10 videos per category from each platform. We selected both the \textit{top five most viewed videos} to capture the most influential content and another \textit{five randomly-sampled videos} to represent a diverse range of content. We then removed videos that were 1) duplicated, 2) non-English, and/or 3) neither about ADHD nor created by ADHD or health-related creators. 

\textbf{\textit{Final Video Dataset.}} The sampling process resulted in 184 TikTok and 189 YouTube videos in the final video dataset with a wide range of views, likes, and comments. On average, videos we selected from YouTube had been viewed $1.7M \pm 3.3M (Mean \pm SD)$ times, and received $77.6K \pm 165K$ likes and $5.6K\pm 24.4K$ comments. Videos from TikTok had been played $4.6M \pm 8.4M$ times, receiving $584K \pm 1.2M$ diggs (likes), and $5.9K \pm 21.1K$ comments.

\subsubsection{Comment Collection.}
Using the same tools for video collection, we collected the top 20 comments under each video on December 11th, 2023. We did not limit comment collection to those explicitly made by viewers with ADHD, as different stakeholders (e.g., health experts, family of people with ADHD) could all play crucial roles in the ADHD online communities \cite{eagle2023}. %\yuhang{did we also collect creator profile and meta data from the video?}

\subsection{Data Analysis}
%\hazel{Not many papers in ICWSM used thematic analysis, but for those that did, some paper presented Cohen's Kappa and some didn't. I think we don't necessarily have to?}
We analyzed the videos and comments with a mixed methods approach, combining both quantitative and qualitative analyses. 

\subsubsection{Quantitative Analysis.}
We compared the characteristics and distributions of ADHD-relevant videos on YouTube and TikTok via Chi-square tests with Bonferroni correction. In addition, we also reported descriptive statistics, including mean, standard deviation, and the proportions of videos with different attributes (e.g., creator types, content types, video forms, different quality control and accessibility measures). 

%We used both descriptive and inferential statistical analysis to quantitatively examine and compare the characteristics and distributions of different types of ADHD-relevant videos on YouTube and TikTok. This included mean value, standard deviation, and the proportions of videos with different attributes (e.g., creator types, content types, video forms, different quality control and accessibility measures). To uncover platform-level differences, we conducted Chi-square tests with Bonferroni correction to investigate whether the distribution of attribute values (e.g., creators self-identifying as having ADHD vs. not, under the attribute \textit{creator type}) differed significantly between the two platforms.
%\yuhang{i think you missed this comment, so highlight here again: need to describe your quantitative method too, are you using descriptive statistic analysis? such as mean, SD? Otherwise how did you get the those quantitative results for distributions, and how can we say that we are using a mixed method?}

\subsubsection{Qualitative analysis.} We employed thematic analysis \cite{Braun2022} to qualitatively generate codebooks and themes to deeply understand the content of ADHD-relevant videos on VSPs and viewers' experiences. 

To analyze the video content and characteristics, we first randomly sampled 20 TikTok videos and 20 YouTube videos from the final video dataset. Two researchers %\yuhang{what other info?profile?description?hashtags?} %After discussing the observations, we agreed to not only watch and transcribe the videos, but also analyze video descriptions and creators' profiles, as they convey important information that could potentially affect the viewers, including creator identity, sources of information, additional resources, and disclaimers. We regard such information as important because they reflect attributes of assessment benchmarks described in Section \ref{informationAssessment} (e.g. creator identity reflects authorship). Beyond the video contents, we also observed and labeled how the content is presented, including video forms and accessibility features such as usage of captions, which could be helpful for neurodiverse users \cite{simpson2023}. %Finally, we decided to have a combined codebook for both TikTok and YouTube videos for more straightforward comparison.
coded the selected videos independently via open coding. To qualitatively analyze the videos, they converted the multimodal video data into text transcripts: First, they obtained the transcript for each video. They then watched the video, took notes of important video content (e.g., presentation form), and inserted the notes into the transcript based on the timestamp. They also added other metadata (e.g., creator profiles) to the transcript, and coded the finalized transcripts following the standard thematic analysis method. After coding the 40 selected videos, they discussed and revised their codes to resolve any disagreement and developed an initial codebook with consensus. Three researchers then divided and coded the remaining videos. During the process, the researchers met and checked their codes on a weekly basis to ensure consistency and add new codes to the codebook. Meanwhile, a fourth researcher oversaw the entire coding process to ensure higher-level agreement. The final codebook contained 61 codes. Two researchers then conducted axial coding together, grouping codes of similar types under high-level categories using affinity diagramming and obtained nine themes. 

We applied the same method to code video comments, with two researchers developing the initial codebook for comments on the 20 sampled videos, and three researchers coding the remaining comments. The final codebook contained over 90 codes with 15 high-level themes. 

Finally, to examine the relationships between videos and comments (e.g., viewers' response to videos with certain quality measures), we conducted another round of axial coding, cross-referencing and connecting codes from the video and comment codebooks, generating nine additional themes. 
%\yuhang{to complete.} 

\subsection{Ethical Considerations}
All videos collected in this work are public. We obtained the Institutional Review Board (IRB) approval before starting data collection. All researchers have prior experience working with individuals with disabilities and/or analyzing data from online communities. Some of the researchers on the team identify as neurodivergent. We recognize the limitations of our experiences and perspectives, and the fact that we, as researchers in HCI, do not have expertise in the health field. The goal of this work is not to judge whether ADHD content on VSPs is high- or low-quality, nor to criticize any content or users. Instead, we examined the voices of the ADHD community members via the videos and the comments to broadcast the underlying issues and needs, thus deriving design implications to support open, safe, and trustworthy content sharing on VSPs.

\section{Findings}
%\yuhang{beginning paragraph...}
In this section, we present the main findings from our mixed methods analysis, including quantitative results on the characteristics of ADHD-relevant videos, the collective efforts by creators and viewers in video quality control alongside their challenges, and ADHD-relevant accessibility issues and practices in video creation.

\subsection{Characterizing ADHD-relevant Videos}
\label{sec:quantitative}
We report quantitative results to \textit{characterize} and \textit{compare} the ADHD-relevant videos on YouTube and TikTok (e.g., creator types, content types, presentation forms), highlighting the commonalities and differences between the two VSPs. 

\begin{figure*}
    \centering
    \includegraphics[width=\linewidth]{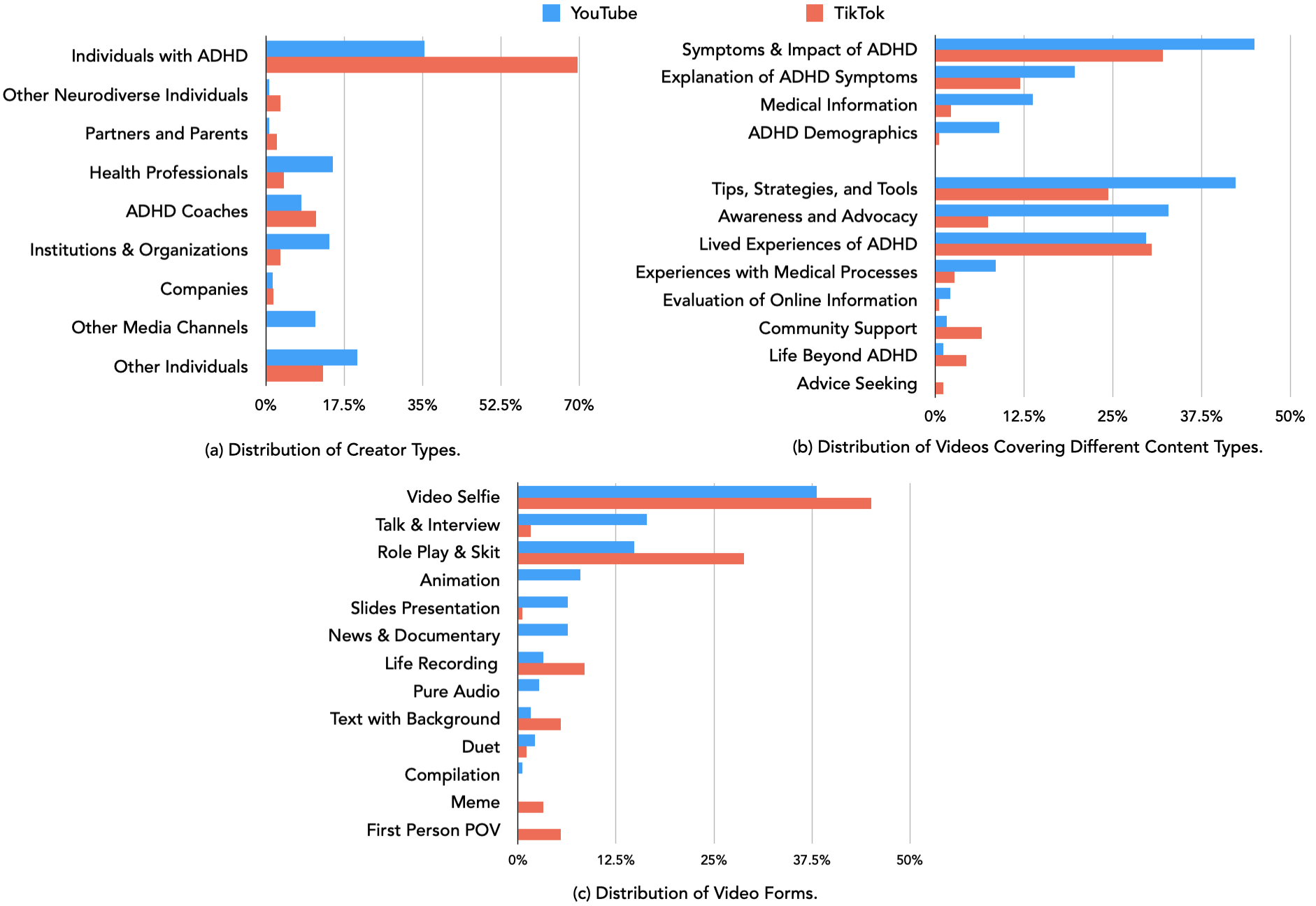}
    \caption{Creator and video distributions on YouTube and TikTok. (a) Distributions of creator types; (b) Distributions of videos convering different contents; (c) Distributions of video forms. 
    }
    \Description{This is a set of three bar charts representing creator and video distributions on YouTube and TikTok. 
    Figure 1a: This bar chart shows the distribution of ADHD-related content creators on YouTube (blue) and TikTok (red). Individuals with ADHD contribute 69.6\% on TikTok and 35.4\% on YouTube. Other creator categories, like health professionals, organizations, and other media channels are more prevalent on YouTube, while TikTok has a higher percentage of other individuals and ADHD coaches.
    Figure 1b: This bar chart compares ADHD video topics on YouTube (blue) and TikTok (red). On YouTube, most content covers ADHD symptoms (44.97\%), strategies (42.33\%), and lived experiences (29.63\%). TikTok content focuses more on symptoms (32.1\%), lived experiences (30.43\%), and community support (6.52\%).
    Figure 1c: This bar chart compared the different forms of ADHD-related videos on YouTube (blue) and TikTok (red). Both platforms share video selfies as the most common format. TikTok features more role plays (28.8\%) and life recording (8.5\%), while YouTube covers more talk and interview (16.4\%) and slides presentation (6.3\%).}
    \label{fig:distributions}
\end{figure*}

\subsubsection{Who are the creators?} In total, we collected videos from 127 creators on YouTube and 125 creators on TikTok. Based on the creator profiles and their self-disclosed identity in the videos, we identified nine types of creators based on their relationship with ADHD, including \textit{individuals with ADHD (self-disclosed)}, \textit{individuals with other neurodiverse conditions} (e.g., autism), \textit{partners and parents of people with ADHD}, \textit{health professionals} (e.g., doctors and psychiatrists who could diagnose ADHD), \textit{ADHD coaches} (i.e., creators who offered paid courses or coaching sessions in a non-clinical setting), \textit{educational institutions and non-profit organizations} (e.g., official channels of universities), \textit{companies that sell ADHD-related products} (e.g., video games for people with ADHD), \textit{other media channels} (e.g., local news), and \textit{other individuals} (i.e., creators who did not disclose any relationship with neurodiversity or health expertise). %These categories are derived based on (1) whether the creator represents an individual or an organization, (2) whether the creator offers services through their account, (3) the health expertise of the creator, and (4) the neurodiverse identity of the creator.  
Note that these categories are not mutually exclusive (e.g., some health professionals and ADHD coaches self-reported as having ADHD). We show the distribution of creator types in Figure \ref{fig:distributions}(a).  

% \begin{figure}
%     \centering
%     \includegraphics[width=\columnwidth]{Figures/creator_types.png}
%     \caption{Creator distributions on YouTube and TikTok. %\yuhang{merge Figure 1-3, so that you create a row with three figures side by side. See the first figure in GazePrompt (Ru's paper); consider increasing the font size proportionally to ensure visibility after reducing the size the figure.}}
%     }
%     \label{fig:dist-creator}
% \end{figure}
% \textbf{\textit{TikTok and YouTube harbor different types of creators of ADHD contents.}} 
Interestingly, we observed a stark difference in the creator type distributions between TikTok and YouTube. 
We found that individuals with ADHD were the dominant creators on TikTok, comprising 69.6\% of all creators producing ADHD-related content, which was significantly more than the 35.4\% on YouTube ($\chi^2(1, 252) = 28.1$, $p < 0.001$). In contrast, YouTube featured significantly more institutions and organizations (YouTube: 14.2\%, TikTok: 3.2\%, $\chi^2(1, 252) = 8.2$, $p = 0.038$), and also had a higher proportion of health professionals (YouTube: 15.0\%, TikTok: 4.0\%), although this difference was not statistically significant ($\chi^2(1, 252) = 7.6$, $p = 0.053$). %We also noticed YouTube to have more health professionals (15.0\%). 
%Moreover, we found that 60\% of the health professional creators on TikTok self-disclosed to have ADHD, while only 10.5\% of those on YouTube did.
The different creator identity distribution indicated a major difference between the two platforms: \textit{TikTok fostered close-knit communities of individuals for personal experience sharing, while YouTube conveyed more ``authoritative information'' from professionals and organizations.} This observation was echoed by viewers' sentiments on the two platforms. For example, a TikTok creator shared their personal experience with adult ADHD diagnosis, encouraging viewers to support each other through this journey. Comments directly responded to the creator's words, highlighting a strong sense of community: \textit{``None of us is alone. You’re building a community of support!'' (TikTok comment).} In contrast, stronger medical expert presence on YouTube allowed viewers to acquire authoritative information, as one viewer wrote under a clinician’s video on non-medical ADHD interventions: \textit{``This channel is real, reliable and absolutely helpful… I’m not able to see a good doctor financially… watching this video really helped'' (YouTube comment).}

\begin{figure*}
    \centering
    \includegraphics[width=0.95\linewidth]{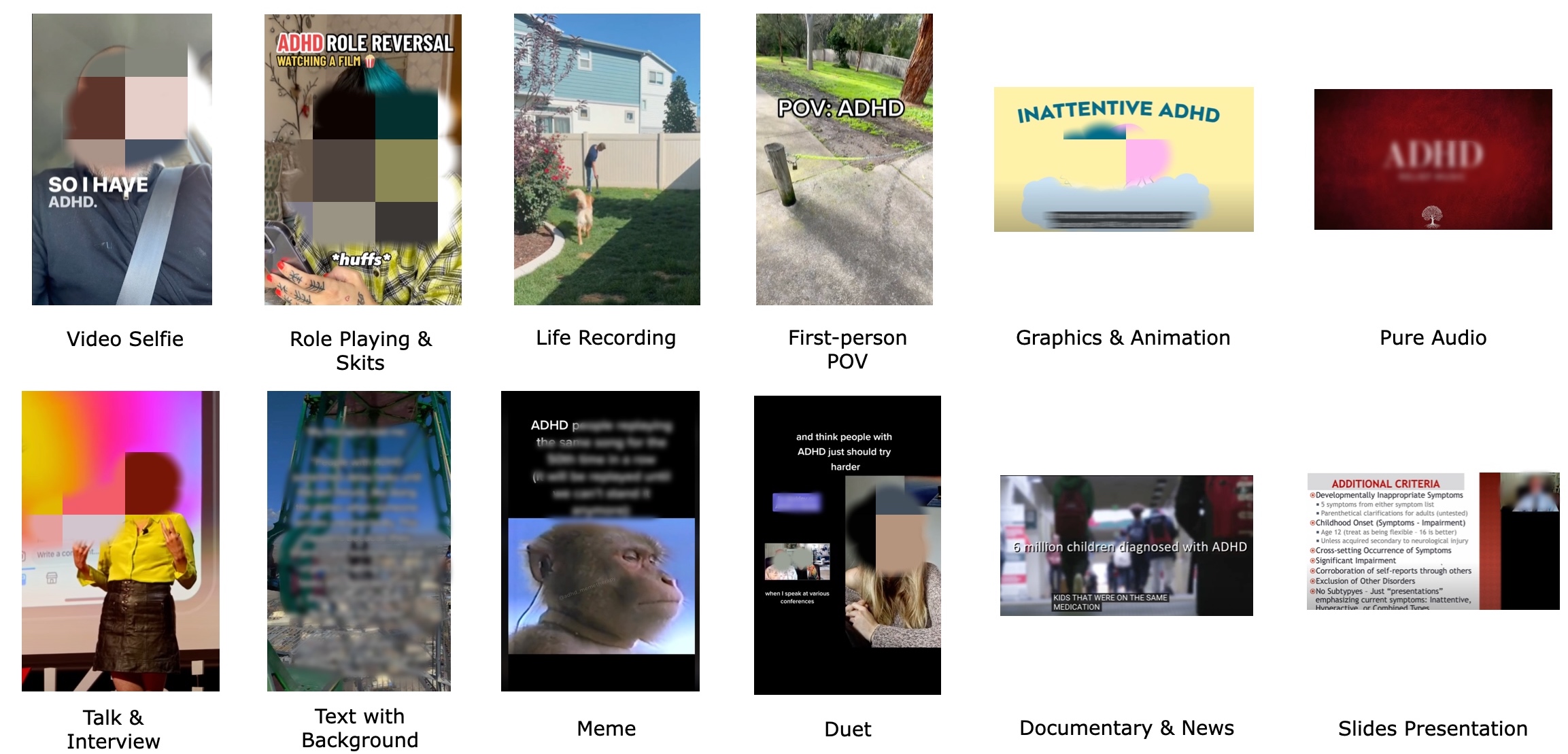}
    \caption{Examples of different videos forms except for compilation. A compilation video could consist of multiple forms above.}
    \Description{This figure showcases examples of twelve different video forms. From left to right and top to bottom, they are: (1) video selfie(i.e. recording of creator speaking to the camera); (2) role playing and skits that vividly illustrated ADHD symptoms and experiences; (3) life recording (e.g., vlogs) that showed creators' own recording of their lives; (4) first-person POV that directly reflected ADHD perspectives; (5) animation; (6) pure audio (e.g., music for helping with ADHD concentration); (7) talks and interviews that reflected public discourse; (8) text with background; (9) memes; (10) duet (i.e., a creator responding to another video); (11) news and documentary that recorded or reported ADHD from a third-party perspective; (12) slides presentation.}
    \label{fig:video-examples}
\end{figure*}

\subsubsection{What contents are covered?}
We identified 12 content types for ADHD-relevant videos. We found that the video content covered two high-level categories---\textit{clinical knowledge \& treatment} and \textit{pragmatic experiences \& strategies}---echoing Zhang et al.'s content taxonomy for mental health videos on TikTok \cite{zhang2023}. We further expanded the prior taxonomy by deriving subcategories specialized for ADHD. Specifically, for the clinical knowledge \& treatment category, we identified four types: (1) \textit{symptoms and impact of ADHD}, (e.g., being inattentive); (2) \textit{explanation of ADHD symptoms} (e.g., using dopamine levels to explain ADHD); (3) \textit{medical information} (e.g., different types of medications); and (4) \textit{demographics of ADHD} (e.g., gender distributions). For the pragmatic experiences and strategies, we identified eight types: (1) \textit{tips, strategies, and tools} (e.g., body doubling); (2) \textit{lived experiences of ADHD} (e.g., growing up with ADHD); (3) \textit{awareness and advocacy} (e.g., explicitly advocating ADHD as a serious disorder); (4) \textit{experience with medical processes} (e.g., clinical diagnostic experiences); (5) \textit{community support} (e.g., play a happy song for \textit{``my adhd people''}); (6) \textit{life beyond ADHD} (e.g., wedding vlogs); (7) \textit{evaluation of online information} (e.g., criticizing certain videos to convey misinformation); and (8) \textit{advice seeking} (e.g., what to do in face of Adderall shortage). We show the distribution of video content on YouTube and TikTok in Figure \ref{fig:distributions}(b) (Note that some videos cover multiple types of content). 

%  \begin{figure}
%     \centering
%     \includegraphics[width=\columnwidth]{Figures/content_types.png}
%     \caption{Content distributions on YouTube and TikTok. %\yuhang{merge Figure 1-3, so that you create a row with three figures side by side. See the first figure in GazePrompt (Ru's paper); consider increasing the font size proportionally to ensure visibility after reducing the size the figure.}}
%     }
%     \label{fig:dist-content}
% \end{figure}

\textbf{\textit{Content Commonalities.}}
We found that ADHD symptoms and impacts were the most common topics on both YouTube (45.0\%) and TikTok (32.1\%). The prevalence of this content corresponded to people's need for self-recognition of ADHD symptoms for further diagnosis or help-seeking. %\yuhang{do you mean self-diagnosis, or simply recognize potential symptoms then seek for formal diagnosis?}. 
To supplement these generic descriptions, many videos on both platforms (YouTube: 29.6\%, TikTok: 30.4\%) shared personal lived experiences with ADHD, allowing viewers to understand ADHD symptoms with concrete examples and contexts. 

\textbf{\textit{Content Differences.}}
 We found that there were significantly more YouTube videos covering medical information (YouTube: 13.8\%, TikTok: 2.2\%, $\chi^2(1,373) = 15.4, p < 0.01$) and ADHD demographics (YouTube: 9.0\%, TikTok: 0.5\%, $\chi^2(1,373) = 12.7, p < 0.01$), which was in line with the larger number of health professionals and organizations on YouTube. YouTube also had significantly more videos on awareness and advocacy, explicitly defying ADHD stereotypes and advocating for ADHD as a serious disorder (YouTube: 32.8\%, TikTok: 7.4\%, $\chi^2(1,373) = 34.9, p < 0.001$), reflecting purposes of many educational videos from organizations. Additionally, although both platforms provided practical tips, strategies, and technologies that facilitated navigation of ADHD lives, YouTube had a significantly higher proportion of videos covering such content (YouTube: 42.3\%, TikTok: 24.3\%, $\chi^2(1,373) = 12.6, p < 0.01$). However, content on TikTok appeared to foster more direct and intimate interactions between ADHD community members, even though the differences were not statistically significant. For example, more videos on TikTok offered direct community support (TikTok: 6.5\%, YouTube: 1.9\%, $\chi^2(1,373) = 4.8, p = 0.36$), and shared personal life events beyond their ADHD experiences (TikTok: 4.4\%, YouTube: 1.1\%, $\chi^2(1,373) = 2.7, p  = 1.00$). As a result, the two platforms could potentially attract different viewers who sought different types of information and support.

\subsubsection{How are videos presented?}

% highlight contribution (explain each form)

% \begin{figure}
%     \centering
%     \includegraphics[width=\columnwidth]{Figures/video_forms.png}
%     \caption{Video form distributions on YouTube and TikTok. %\yuhang{merge Figure 1-3, so that you create a row with three figures side by side. See the first figure in GazePrompt (Ru's paper); consider increasing the font size proportionally to ensure visibility after reducing the size the figure.}}
%     }
%     \label{fig:dist-forms}
% \end{figure}

% \begin{figure*}
%     \centering
%     \includegraphics[width=0.95\linewidth]{Sections/Figures/video_forms_distribution.png}
%     \caption{Distributions of different video forms on YouTube and TikTok. }
%     \label{fig:video-forms}
% \end{figure*}

We found that creators shared ADHD-related videos via various presentation forms based on their purposes. We identified 13 forms of video presentations on YouTube and TikTok (Figure \ref{fig:video-examples}), including (1) \textit{video selfie} (i.e., recording of creator speaking to the camera) %\yuhang{need explanation, the same to the rest categories}
; (2) \textit{role playing and skits} that vividly illustrated ADHD symptoms and experiences; (3) \textit{life recording} (e.g., vlogs); (4) \textit{first-person POV} that directly reflected ADHD perspectives; (5) \textit{animation}; (6) \textit{pure audio} (e.g., music to help with ADHD concentration); (7) \textit{talks and interviews} that reflected public discourse; (8) \textit{text with background}; (9) \textit{memes}; (10) \textit{duet} (i.e., a creator responding to another video); (11) \textit{news and documentary} that recorded or reported ADHD from a third-party perspective; (12) \textit{slides presentation}; and (13) \textit{compilation} (i.e., stitching multiple videos together to create a new one). We show the distribution of video presentation forms in Figure \ref{fig:distributions}(c). 

%\textbf{\textit{TikTok and YouTube have distinct styles of presenting ADHD contents.}}
Our results showed that the most common presentation form on both YouTube and TikTok was a video selfie (YouTube: 38.1\%, TikTok: 45.1\%). However, the two VSPs also leveraged different presentation forms for ADHD content. Specifically, YouTube had significantly more \textit{formal} videos presented from public perspectives, including talks and interviews (YouTube: 16.4\%, TikTok: 1.6\%, $\chi^2(1,373) = 22.8, p < 0.001$) and news and documentary (YouTube: 6.3\%, TikTok: none, $\chi^2(1,373) = 10.1, p = 0.019$). YouTube also uniquely presented animation, pure audio, and compilations, reflecting production effort and diversity of presentation forms in long videos. In contrast, TikTok had significantly more videos featuring role-playing (TikTok: 28.8\%, YouTube: 14.8\%, $\chi^2(1,373) = 9.9, p = 0.021$), and uniquely presented memes and first-person POV videos. Video forms on TikTok illustrated ADHD experiences in a \textit{personal, creative, and light-hearted} manner, reflecting different styles and community cultures between TikTok and YouTube.
\subsection{Efforts and Challenges in Content Quality Control}
We elaborate on the collective efforts---by VSPs, creators, and viewers---in indicating, assessing, and improving the quality of ADHD-relevant videos, alongside the underlying challenges. While some findings might generalize to other health communities, we highlight the unique interactions and perceptual gaps across different user groups in ADHD communities on VSPs, which were critical in people's help-seeking decisions due to the long-running stereotypes of ADHD and the diagnostic challenges that followed \cite{eagle2023}.

\subsubsection{Proactive Quality Indicators on VSPs.}

\label{sub-sub-sec: credibility-practices}

We found that both VSPs and creators proactively indicated video qualities via authorship, attributions, and areas of uncertainty. We elaborate on their strategies and the corresponding challenges. 

\textbf{\textit{Establishing Authorship: Identity Disclosure by Creators.}} Authorship, including authors, their affiliations, and relevant credentials, plays a vital role in trust building \cite{Silberg1997}. We found that creators of ADHD-relevant videos on both YouTube and TikTok disclosed their \textit{neurodiverse conditions} and \textit{ADHD- or health-related expertise} to provide authorship information, thus establishing authority and trust. Creators disclosed their identities via profiles, video content, video descriptions, creator names, and video titles, with most ADHD individuals (YouTube:
55.6\%, TikTok: 69.5\%) and health professionals (YouTube: 78.9\%, TikTok: 100\%) disclosing their identities via profiles.
%\added{We noticed that creators disclosed their identity in a variety of places.} Creators with ADHD most commonly disclosed their neurodiverse identity via their profile (YouTube: 55.6\%, TikTok: 69.5\%), followed by self-disclosure in videos (YouTube: 44.7\%, TikTok: 39.4\%), video descriptions (YouTube: 20.5\%, TikTok: 6.6\%), and video titles (YouTube: 9.2\%, TikTok: none). Health professionals also most commonly used profiles to disclose their expertise (YouTube: 78.9\%, TikTok: 100\%), followed by keywords (e.g., Dr, Ph.D.) in creator/channel names (YouTube: 52.6\%, TikTok: 40\%), self-disclosure in videos (YouTube: 38.5\%, TikTok: 40\%), video descriptions (YouTube: 26.9\%, TikTok: none), and video titles (YouTube: 7.7\%, TikTok: none). \yuhang{this paragraph is less interesting...a potential place to trim to one sentence and merge to last paragraph, removing all the stats} 

\textbf{\textit{Challenge: Vague Disclosure.}} %\added{As a neurodeverlopmental disorder with high rates of comorbidity, ADHD diagnosis and treatment could involve a variety of professionals \cite{CHADD2024}, making it challenging to identify the proper help.} 
The various authorship establishment strategies also raised issues. We found that creators lacked a common standard for identity and background disclosure, with creators sharing their identities to varying degrees, potentially leading to misleading information. For instance, 8.3\% of YouTube and 60\% of TikTok creators who claimed to be ``Dr.'' or ``Ph.D.'' provided only degree information with no further expertise details (e.g., clinical psychologist). This lack of specificity could sometimes lead to confusion or misinterpretation. For instance, one TikTok creator disclosed their Ph.D. degree in their profile with no expertise specification; however, a website linked from their profile indicated that the Ph.D. degree was not related to health or ADHD. This incomplete disclosure on TikTok resulted in confusion among viewers as reflected in the comments:
\begin{quote}
\textit{Viewer A: ``Can you share the rationale behind your statement?'' (TikTok comment).}

\textit{Viewer B: ``Their PhD'' (Reply to Viewer A).}

\textit{Viewer C: ``They have a Ph.D. in [a communication-related field]'' (Reply to Viewer B).}
\end{quote}

%a TikTok comment asked if the creator could share the sources behind their video contents, to which another comment responded: \textit{``[Their] PhD.''} The creator disclosed this identity directly on TikTok. However, the first viewer followed up: \textit{``[They] have a Ph.D. in [a communication-related field]''}, 
This conversation indicated viewers' skepticism towards the creator's authorship. %The specific degree information was not disclosed on the creator's TikTok account but can be found on a website linked in their profile. 
While the creator might not intend to associate their Ph.D. with the reliability of video content, some viewers could misinterpret such information due to the incomplete disclosure at important locations (e.g., profiles), resulting in issues such as over-estimating the credibility of the content due to the creator's disclosed authorship.

% \textbf{\textit{Establishing Authorship: Platform Recognition.}} Besides creators, we found that YouTube also adopted a platform-based mechanism to recognize creators' health expertise and label that on videos for viewers. We found that 7.1\% of creators on YouTube (10.6\% of videos) got a one-liner description---\textit{``From a channel with a health professional licensed in [country].''}---to recognize their reliability as health sources. We did not observe a similar feature on TikTok for health sources recognition. %Such an approach acts an external validation for creators' expertise, lending more weights to the reliability of creators' authorship. 

% \added{\textbf{\textit{Challenge: Incomplete Capture.}}} We noticed that this approach might not fully capture the available professional healthcare sources on YouTube, with 52.6\% of creators who self-disclosed as health professionals not being recognized as health sources by YouTube. \yuhang{findings relevant to platform recognition could be removed... not much interesting data}

\textbf{\textit{Establishing Authorship: Platform Recognition.}} Besides creators, we found that VSPs also had platform-based mechanisms to recognize creators' health expertise. Specifically, 7.1\% of creators on YouTube (10.6\% of videos) had a brief description under their videos---\textit{``From a channel with a health professional licensed in [country].''}---to recognize their reliability as health sources. %Such an approach acts an external validation for creators' expertise, lending more weights to the reliability of creators' authorship. 

\textbf{\textit{Challenge: Low Recognition Rate.}}
Despite the effort, we noticed that such a platform-based approach might not fully capture the available professional healthcare sources: the YouTube health expert recognition feature only covered 47.4\% of creators who self-disclosed as health professionals, potentially limiting viewers’ access to some credible medical information.
%Furthermore, we did not find having such a recognition increase video views. On average, these videos receive 499351 $\pm$  904579 views, which is below the average number of views (16666368 $\pm$ 3009435) for all ADHD-related videos on YouTube, and slightly below the average number of views for videos made by health professionals (654482 $\pm$ 1253643). They also receive fewer number of likes and comments than the average. 
At the time of our work, we did not find a similar feature on TikTok for recognizing health sources.

\textbf{\textit{Making Attribution: References.}} Aside from establishing authorship, we also observed some creators added references to their videos for reliable attribution. Common references included health-related publications (e.g., JAMA) and articles by authoritative ADHD-related organizations (e.g., CHADD). %\yuhang{is the following data specific for YouTube?} 
On YouTube, we found that 13.2\% ADHD-relevant videos contained some form of references: 7.9\% in video descriptions, 4.2\% in video contents, and 2.6\% vaguely mentioning some studies without explicitly providing titles or links. Notably, one video added references with timestamps in the video description to help viewers map references to the video content. 
Compared to YouTube (13.2\%), significantly fewer videos on TikTok (2.2\%) included references ($\chi^2(1, 373) = 14.4, p < 0.001$), with 0.55\% of videos having references in descriptions and 1.65\% in videos. 

We found that some viewers would actively seek references from non-professional creators to ensure the quality of the ADHD content they consume. For example, under a YouTube video that promoted a video game claimed to alleviate ADHD symptoms, a comment pointed out the video lacked concrete references: \textit{``Where's the studies backing up the bold claim that it improves our symptoms...? Link to peer-reviewed research papers, please?'' (YouTube comment).}

%Moreover, the references provided by TikTok videos tended to deliver fewer details than YouTube videos. For example, 66.7\% videos with in-video references presented them in the form of \textit{``[last name] et al. [year]''} without providing additional information or links, while YouTube videos usually provided links or complete titles of papers or articles. 
%\yuhang{what about youtube videos?}. 
%Such a succinct form of reference presentation on TikTok, while possibly more suitable for the vertical short-video form that has less visual space, might potentially cause difficulty for viewers to identify the study sources and further raise quality and trust issues. 

\textbf{\textit{Challenge: Irrelevant References.}}
We found some creators added irrelevant references to their videos to make the content appear credible. For example, one video on TikTok about ADHD behaviors shared in its description that \textit{``This video is based on the following scientific article: [link]'' (TikTok description).} However, %multiple comments showed disagreement towards this video: \textit{``This is too broad. Could happen to anyone. How is this [about] ADHD?''}
we reviewed the reference link, but found no content relevant to ADHD or neurodiversity in it. Comments further highlighted concerns about the misconception introduced by the video: \textit{``Videos like this make people think they have ADHD, and ended up with more stigma on ADHD'' (TikTok comment).} Adding irrelevant references could bring the risk of misleading viewers into trusting low-quality content on a surface level and associating certain behaviors with ADHD without a deep understanding.

\textbf{\textit{Clarifying Areas of Uncertainty: Reminders and Disclaimers.}} In contrast to indicating the high quality of their videos, some creators highlighted uncertainties and limitations of their videos to avoid misinformation. They emphasized, for instance, that they were not health professionals, that the content reflected only personal experiences or opinions, or that the viewers should seek professional help or diagnosis. For example, a TikTok video demonstrating the creator's ADHD symptoms had a reminder in the description: \textit{``Please take this video with humour---it's how I cope with my messed up life'' (TikTok description).} Some creators also added in-video captions reminding viewers that the video is \textit{``not a diagnostic tool'' (TikTok video).}

In total, we found 3.8\% of videos on TikTok included such reminders (2.2\% appearing in videos, 1.1\% in comments, and 0.5\% in descriptions) compared to 15.3\% of videos on YouTube (11.1\% in descriptions, 4.2\% in videos, 2.6\% in profiles, and 0.5\% in comments), with YouTube having a significantly higher proportion ($\chi^2(1, 373) = 12.9, p < 0.001$). These reminders highlighted limitations of the video content, encouraging viewers to be cautious in help-seeking online and acquire clinical help if possible. 

\textbf{\textit{Challenge: Low Visibility.}} Despite creators' efforts, these reminders might not be obvious or visible enough to the viewers. For example, we found 
%an ASMR video (i.e., videos that use sounds or visuals to create a calming, tingling sensation) 
a video on TikTok featuring the creator playing the role of a doctor diagnosing viewers with a series of ``ADHD tests''. The creator put a reminder in the comment: \textit{``I'm not a doctor, and this video is only for your relaxation'' (TikTok comment).} However, there were still viewers seeking professional advice in comments: \textit{``Does it mean I have ADHD if I failed the test?'' (TikTok comment).} Low visibility of these reminders or disclaimers can prevent the creators from effectively conveying their intent to viewers, leading to unnecessary expectations and misunderstanding.

\subsubsection{Quality Control via Comments.}

Besides VSPs and creators, we also found viewers playing an active role in ADHD content quality control via comments---a critical space for ADHD viewers. We elaborate on viewers' strategies for assessing and improving ADHD content quality.

\textbf{\textit{Active Engagement in Comments as Viewers with ADHD.}} We found that comment-reading during videos was a common behavior for viewers with ADHD, especially under long videos on YouTube, due to difficulty sustaining attention being a common ADHD symptom \cite{Wilens2010-hq}. We discovered 30 comments under 19 YouTube videos mentioned finding themselves reading comments halfway through a video. Twenty comments explicitly linked this behavior with their ADHD identity: \textit{``Anyone with ADHD knows that during a video they scroll to the comments'' (YouTube comment).} One comment detailed this experience: 

\begin{quote}
    \textit{``ADHD is like starting to scroll through the comments, constantly paying attention while also reading comments, and then a comment says something interesting, and you realize the video is over and [you] literally have no idea what the video is about'' (YouTube comment).}
\end{quote}

% \deleted{In contrast, we noticed comments on TikTok did not highlight the behavior of comment scrolling as frequently, potentially because TikTok videos were often much shorter and could more easily maintain viewers' attention throughout. \deleted{We only observed similar sentiments when one creator intentionally made a part of the video speechless to \textit{``test the viewers' attention span,''} under which multiple comments mentioned they were \textit{``scrolling through comments to pass the time.''}} Nonetheless, we still found 11 videos on TikTok having comments explicitly underlining the importance of comments for community building, such as \textit{``raise your hand if you found your people in this comment section!''}}

\textbf{\textit{Comments Assessing ADHD Content Quality.}} Given the importance of comments for viewers with ADHD, comments also served as an important space for viewers to assess the video quality. For example, comments under 13 YouTube and 28 TikTok videos highlighted how the video content differed from their own experiences: \textit{``That's far from what ADHD is! I suffer daily and it ain't anything like what you have recorded!'' (TikTok comment)}. Such comments emphasized the limited (and sometimes misleading) perspective of one's personal experience, reminding viewers not to hastily associate a particular behavior with ADHD. In other cases, comments highlighted factual mistakes in six YouTube and three TikTok videos: \textit{``As someone with severe ADHD, this video is NOT an ADHD test. ADHD is NOT getting mesmerized by cool patterns and moving colours'' (YouTube comment).} Such comments served as important signifiers of potential misleading content, making viewers more discerning about the video content.

Despite the critical role of comments in video quality assessment, we recognized that comments could also be a source of misinformation. For example, some comments expressed ADHD  misconceptions under videos sharing personal ADHD experiences: \textit{``This is absolutely bullshit. Stop excusing your laziness!'' (TikTok comment).} %Such comments reflected social misconceptions of ADHD, underscoring the risks of blindly trusting information in comments.

\textbf{\textit{Comments Supplementing Video Contents.}} Aside from showing agreement or disagreement, some comments also made efforts to \textbf{improve} the video content quality. In particular, diverse ADHD experience sharing via comments helped construct a more holistic portrayal of ADHD.  We found comments under 18.0\% of YouTube and 21.9\% of TikTok videos shared alternative or additional perspectives that the videos overlooked. For example, under a YouTube video where the creator described her ADHD experiences as having a \textit{``superpower'' (YouTube video),} a comment highlighted the more challenging aspects of ADHD for many individuals:
\begin{quote}
    \textit{``ADHD IS a disability... I’m happy that [the creator] seems happy with their ADHD, but it’s way worse for some people, and we have a much tougher time'' (YouTube comment).}
\end{quote}

\subsubsection{Perceptual Gaps between ADHD Content Creators and Viewers.} While we found that content creators and viewers collectively contribute to signaling and improving the quality of ADHD-relevant videos, we also recognized some unique perceptual gaps between them. Specifically, some creators with high credibility could be perceived as arrogant by viewers if they lacked ADHD community understanding. In addition, some quality indicators proposed by creators might be impractical to the viewers due to the lack of pointers (e.g., links) to professional resource on VSPs.

\textbf{\textit{Authorities vs. Individuals.}}
%\hazel{tie to adhd stereotypes}
Although prior work \cite{Yeung_Ng_Abi-Jaoude_2022} has regarded health professional creators as high-quality sources of content, we noticed some viewers directly criticized ADHD conceptualization content (e.g., videos on \textit{``What is ADHD?''}) with reliable authorship (e.g., health professionals) as \textit{``arrogant''} and \textit{``dismissive'' (YouTube comment)}. Such criticism was rooted in the conventional portrayal of ADHD from an authoritative lens, framing ADHD as a defect instead of a difference and disregarding the validity of many coping strategies adopted by individuals \cite{Spiel2022}. For example, a comment on YouTube criticized a health professional for their negative attitude towards the use of stimulants by some individuals with ADHD:

\begin{quote}
    \textit{``The way you presented this is atrocious. You focused really heavily on the negative aspects, speaking nothing of the positive affects that give many of us the only chance to function within normal society... We're treated like drug addicts for having ADHD'' (YouTube comment).}
\end{quote}

Perspective differences could lead to a gap in community culture and understanding between the ADHD viewers and the professional creators. Such a gap could reduce the usefulness of the creators' advice, % For example, we found a YouTube video of a talk on the comorbidity between ADHD and autism, where a health professional shared some online resources. Viewers pointed out that certain resources included are inappropriate: \textit{``I find it horrendous that [resource name] was included. It’s a hate group.''} The lack of community understanding could also 
affecting viewers' general impression of the video and its creator: \textit{``I don’t want to hear one word that they have to say''  (YouTube comment).}

\textbf{\textit{Diagnostic Needs vs. Missing Resource Pointers.}}
%\hazel{highlight adhd diagnostic difficulties}
As previously highlighted, many creators reminded and encouraged viewers to seek clinical evaluations of ADHD. However, we found creators rarely offered direct pointers (e.g., links to directories) or guidance on how to access professional resources. In total, we only identified four health professionals on YouTube who included links to their clinics' websites in their profiles or video descriptions, covering merely 2.1\% of videos. We did not observe any creator-initiated pointers to professional resources on TikTok. Additionally, neither of the VSPs offered platform-based resource pointers.

As a result, we found many viewers---especially adults and women who have been historically marginalized in ADHD---underscored the lack of access to \textbf{\textit{good}} clinical diagnostic resources in comments. We identified comments on 13 YouTube and nine TikTok videos mentioning the difficulty of finding a healthcare professional that \textit{``actually understands'' (TikTok comment)}. One YouTube comment specifically complained about the difficulty for female adults to get an ADHD diagnosis: \textit{``I 100\% CANNOT find help or anyone who can diagnose female adults [for ADHD]... Everywhere I call are not able to help'' (YouTube comment).} 

The gap between an encouragement to seek professional diagnosis and a lack of professional resource pointers could make it challenging for people with diagnostic needs to move forward---a problem especially critical for the ADHD community due to the diagnostic challenges faced by the adult and female members. %However, we found that \textbf{little resources shared addressed the critical issue of lack of clinical diagnostic resources.} 
Our findings highlight an unmet need for clinical diagnosis resource pointers on video platforms, which could play a crucial role in guiding viewers through the help-seeking process.

\subsubsection{Tensions around ADHD Self-diagnosis.}

Due to the lack of accessible clinical resources, some viewers had to self-diagnose as having ADHD. However, we found ADHD self-diagnosis to be a particularly controversial topic in the ADHD community on VSPs. We found that some viewers expressed reservation towards the idea of self-diagnosis, with comments expressing frustration over people who irresponsibly claimed to have ADHD and conveyed inaccurate ADHD impressions. For example, multiple comments under a YouTube video discussing video influence on mental disorders shared their encounters: \textit{``[They] claimed to self-diagnose ADHD and would rub it in everyone’s face... Claiming many insulting stereotypes about ADHD were true'' (YouTube comment).} 

We recognized that such a reserved attitude towards ADHD self-diagnosis could help viewers become more aware of the risks associated with hasty self-diagnosis. For example, we found comments on six YouTube and two TikTok videos stated \textit{``I don't want to self-diagnose''} before sharing how they related to the shared ADHD experiences in the videos, and two YouTube comments explicitly expressed that they would never self-diagnose as it \textit{``puts [them] in danger.''} However, the negative perception could simultaneously induce stigma on people who chose to self-diagnose after careful research: \textit{``Unfortunately, I have to self-diagnose myself with ADHD because I couldn’t get help... People wouldn’t take me seriously if I come out and say, `I’m self-diagnosed.''' (YouTube comment).}

Despite the controversy, we also found VSPs offered a space for self-diagnosed individuals to advocate for this diagnostic approach. For example, one YouTube comment shared their improved self-understanding with self-diagnosis: \textit{``I’ve done a lot of research for my self-diagnosis and I'm happy with where I am now. Getting an official one won’t really benefit me, but understanding myself does!'' (YouTube comment).} VSPs played a critical role in sparking viewers' self-recognition and guiding further exploration of ADHD, with 19.0\% of YouTube and 22.8\% of TikTok
videos containing comments that explicitly questioned if they might have ADHD after watching those videos: \textit{``The more I watch these videos, the more convinced I become that I might have ADHD'' (YouTube comment).} Given the need for ADHD self-exploration and the significance of VSPs during this process, our findings underlined the importance of improving content quality control practices to help viewers make more informed decisions when navigating ADHD content.

\subsection{Video Accessibility for ADHD as a Critical Factor Affecting Video Quality}
\label{sub-sec: accessibility}

Given the unique richness and engaging formats of videos, video delivery and accessibility become another critical aspect that affects video quality \cite{Shoemaker2014}, impacting ADHD viewers' ability in fully taking in the video content. %We identified the current video accessibility issues as well as existing efforts by VSPs, creators, and viewers to alleviate these challenges.
%\subsubsection{What makes videos not ADHD-friendly?} 
We identify four primary aspects that impact video accessibility for viewers with ADHD, including prolonged video length, slow pace, distracting sounds and visuals, and missing or low-quality captions. We explicate each aspect below, alongside the current practices and barriers.   

\subsubsection{Video Length.} We found comments under 17 YouTube and two TikTok videos criticized the videos for being too long for viewers with ADHD. Compared to the length of the TikTok videos (1.0 $\pm$ 1.2 minutes) we collected, long videos on YouTube (13.4 $\pm$ 22.3 minutes) were much harder to interpret for viewers with ADHD. For example, under a 2-hour YouTube video on medical information, 40\% of comments complained about its length: \textit{``Isn't it ironic? This will take me a month to get through [this video] with my ADHD'' (YouTube comment).} Some comments highlighted their preference for short videos under a minute: \textit{``Videos for people with ADHD that are 8 minutes long... I am not sitting through them! You helped me so much with these short videos'' (YouTube comment).} 
Among the 17 YouTube videos longer than 30 minutes, 58.8\% were talks or presentations from health professionals, making such professional content especially inaccessible to people with ADHD.

To make the long videos more accessible, both creators and viewers made efforts by leveraging platform-enabled features or manually breaking down the video content. 

\textbf{\textit{Video Chapters.}} We found 27.0\% of videos on YouTube used its \textit{video chapter} feature (11.6\% of videos relied on automatically generating the chapters), providing a ``table of contents'' with section titles and timestamps for easier in-video navigation. A comment indicated appreciation for this feature: \textit{``No way we can finish a video this long... When you have the chapters, it's easier'' (YouTube comment).}
However, organizing videos into chapters could be challenging for creators with ADHD. We discovered one YouTube video where a creator showcased multiple ADHD-relevant products and promised to create chapters based on products. However, only two products were properly labeled, despite the video featuring over a dozen. Comments attributed this incomplete effort to the creator’s ADHD, with one stating: \textit{``This is the most ADHD thing ever!'' (YouTube comment).}

\textbf{\textit{Community-driven Video Breakdown.}} Besides the platform-enabled feature, %Although video chapters facilitated video navigation, they didn't alter the video length. \deleted{We still found 58.8\% of videos with video chapters having comments criticizing their lengths.  %\yuhang{i did not understand...does video chapter change the original video length? need more explanation}. %One potential reason that contributes to the persisting challenge is the lack of informativeness of the video chapters. Besides the limited length of chapter titles and the efficacy of automatically chaptering videos, creators might also hold different perspectives than their viewers in manually editing video chapters, making those chapters less helpful than they intended to be. 
%These persisting complaints indicated that some video chapters provided were not as useful for the viewers.} To tackle this challenge, 
we also found comments under 12 different YouTube videos contributed their own breakdowns of the video, sharing personal notes followed by timestamps directly linked to the videos. To reduce the amount of content to watch, some comments focused only on key content, such as \textit{``actionable items in video'' (YouTube comment).} %while others cover the entirety of the video content. 
These efforts highlighted the value of a community-driven approach in making the video content more accessible.
 %Considering the prevalence of neurodiverse creators making ADHD-relevant content (Section \ref{subs-sub-sec:creator-type}), extra efforts were required for them to make their videos ADHD-accessible. 

% Similar sentiments could also occur for relatively shorter videos. For example, Under a 20-minute YouTube video that involves an interview with a health professional, one viewer commented:

% \begin{quote}
%     \textit{``[You] really need to cater to your audience. Not saying this to be rude, but I came for short and sweet information. Instead I got a lot of information, and I couldn't hear [it well] because of ADHD.''} 
% \end{quote}

% \deleted{We also recognized that viewers of ADHD-relevant videos could have different thresholds of comfortable video lengths. For instance, even though multiple viewers had trouble watching a 2-hour video, we could also see  comments under the same video shared: \textit{``Although I have ADHD, I enjoyed watching this video and dedicated my full capacity of attention to learn.''} In contrast, we identified comments on TikTok complaining that \textit{``5 minutes is a long time.''} We also found some comments that explicitly clarified their preference for under-a-minute short videos on YouTube: \textit{``I’ve seen videos for people with ADHD that are 8 minutes long... I am not sitting through them! You helped me so much with these short videos.''}}

\subsubsection{Distracting Sounds and Visuals.} Comments under seven YouTube and three TikTok videos complained about distracting auditory or visual elements. For example, under a documentary on YouTube that depicted the lives of families affected by ADHD, a comment described the soundtracks behind certain scenes (consisting of multiple sounds overlaid on each other) as \textit{``maddening for someone with ADHD'' (YouTube comment).} Besides audio, some comments also expressed a preference for a more simplistic visual design. For example, under a YouTube video illustrating ADHD experiences via anime clips, a comment criticized the video design: \textit{``As a person with ADHD, the anime overlays are too distracting. I would rather see [the creator's] face or mouth speech patterns because it helps me focus'' (YouTube comment).}

% \deleted{Similarly, the physical environment where a creator filmed themselves could also become distracting, as a viewer noted under a role-playing video on TikTok: \textit{``I couldn't focus. The bunny [in the background] was too cute to be ignored.''}}

Other than elements in videos, viewers also complained about being distracted by VSP features such as video recommendations: \textit{``It's like a disease'' (YouTube comment)}. One YouTube comment described their desire for an \textit{``ADHD mode''} interface, with \textit{``only a clean screen with the specific content you’re viewing and no recommended videos'' (YouTube comment).}

\subsubsection{Slow Pace.} Video pace (i.e., how fast a creator speaks and how long it takes for them to \textit{``get to the point''}) was shown to be another factor that could cause accessibility issues for viewers with ADHD. Six YouTube and one TikTok videos received comments complaining about the slow pace of the video. As a comment mentioned: \textit{``An ADHD symptom for me: I struggle with people who talk slowly... I lose focus quickly if information isn’t given to me at lightning speed'' (YouTube comment).}

Interestingly, we discovered that viewers with ADHD could perceive \textit{``fast''} differently from neurotypical people. For example, we found that a health professional giving a talk apologized for speaking too fast. A comment responded to this apology: \textit{``[He] made me laugh when he apologized for speaking fast, since I’m watching this at 1.5x speed. [I] wouldn’t be able to focus or process at a lower speed'' (YouTube comment).} 

To address this issue, we found that comments under 14 YouTube and three TikTok videos suggested speeding up the videos to make them easier to watch. % such as \textit{``People with ADHD: Play this one 1.5x speed if you keep losing focus'' (YouTube).} %However, we also found some comments might convey misleading information that directly relates ADHD with the speed adjustment strategy: \textit{``If you put this video on 2x, you definitely have ADHD.'' (TikTok)}
%Despite the common usage, we found that the speed adjustment feature is not always evident to users on TikTok. We identified one TikTok video where an individual with ADHD explicitly explained this feature (i.e. long press either side of the screen to speed up a video) to the users. Multiple comments showed appreciation towards this tutorial: \textit{``Thank you for explaining how the 2x function works. I’ve only done it accidentally and haven’t been able to figure out how it happened.''}
However, we found that speeding up a video could sometimes cause unintended side effects. As most videos leveraged a multimodal format, speeding up a video would also increase the playback rate of visual information, which may cause further distraction for people with ADHD. We found a YouTube comment complaining about the distracting effect of speeding up visual information: \textit{``I turned up the speed to 1.5x, but then [the creator's] blinking became weird and I focused on that. I had to replay the video all over'' (YouTube comment).}

%\deleted{In contrast, we found viewers explicitly appreciated video creators for their fast speaking speed. For example, under a YouTube talk given by a health professional, a viewer commented: \textit{``Finally someone explaining ADHD fast enough so that people with ADHD [could] actually keep listening.''}} % to which another viewer echoed: \textit{``He talks so fast [that] I was actually able to listen to 90\% of the talk!''}. 

\subsubsection{Captions \& Subtitles.} Echoing prior work \cite{simpson2023}, we found comments under one YouTube and three TikTok videos expressing a preference for captions and subtitles. %One video on TikTok specifically discussed their preference for subtitles, sharing the belief that having subtitles could help people with ADHD stay engaged. 
A comment explained this need with their challenges in processing speech: \textit{``I won’t be able to understand if there aren’t subtitles. I speak English, but I can barely hear English---if that makes sense'' (TikTok comment).} 

%We found 99.5\% videos on YouTube had captions (75.7\% auto-generated). %The auto-captioning functionality was not available on one video created by a health professional 9 years ago. 
%On TikTok, we found 30.4\% videos with creators offering their own speech captioning, 35.3\% videos providing textual descriptions to offer additional comments or capture the essence of a scenario (e.g., \textit{``house cleaning day!''}), and 34.1\% videos relying on auto-captions, which is 
%only available for the mobile users.
While current VSPs commonly support auto-generated captions, we found that caption quality was especially important for viewers with ADHD. One comment shared how incorrect subtitles could be distracting: \textit{``It drives me insane when [subtitles] are wrong... My ADHD brain will just hyperfocus on spotting every single mistake'' (TikTok comment).}

%Despite the wish, we did not find missing captions to be a prominent issue, potentially because both platforms offer auto captioning. %Across all videos on YouTube and TikTok, we only found one comment under an 1-hour slides presentation on YouTube explicitly mentioned: \textit{``I really wish the video was subtitled.''} 
% We detail current subtitle practices on YouTube and TikTok in the next subsection.

%\subsubsection{Accessibility Practices and Barriers.} 
%We found that both creators and viewers adopted VSP features and practices to improve video accessibility for viewers with ADHD. We detailed these strategies and their issues below. % accessibility issues for ADHD, With respect to the accessibility issues above, we examined platform features and user practices that might alleviate those issues, even though such features are not necessarily designed for ADHD accessibility. We analyzed their usefulness and sufficiency as reflected in comments.  

%\textbf{\textit{Captions and Subtitles.}} 
\section{Discussion}
Our research has contributed an in-depth examination of long and short-form ADHD-relevant videos across \textit{different video platforms}, uncovering the video quality issues, user strategies, and challenges via a mixed-method approach. We answered the three research questions proposed in the Introduction: 

For RQ1, we characterized ADHD-relevant videos on YouTube and TikTok. We uncovered that the two platforms, due to their different video characteristics, served different purposes but also posed different video quality and accessibility issues. For example, longer videos on YouTube, often made by health professionals, could pose accessibility challenges for ADHD viewers due to length and pace. In contrast, TikTok videos, with their personal and humorous styles, rarely contain quality control measures, which can lead to limited or sometimes misleading perspectives.
%\yuhang{explain how different platform characteristics may lead to different issues.... the prior version is too vague without any specific findings. See my example but expand.}

RQ2 focused on the creators' aspect, where we examined their efforts in proactively indicating content quality through identity disclosure, attributions, and reminders/disclaimers, alongside the corresponding issues underlying such strategies (e.g., incomplete identity disclosure). We also revealed videos' accessibility issues and creators' practices and challenges in improving ADHD accessibility (e.g., difficulty of creating video chapters as an ADHD creator).

For RQ3, we focused on the viewers' perspective via comment analysis, uncovering their reactions to quality and accessibility issues as well as their strategies and challenges in assessing and combating these issues as a community, such as questioning the authoritative sources, providing supplementary information to the video content, and commenting with video summary and timestamps to facilitate accessibility. % For example, xxx \yuhang{complete}

\subsection{Video Quality Control in the ADHD Context} 
Our research confirmed that content quality control on social media is a collective effort across multiple stakeholders \cite{agichtein2008finding}, including social media platforms \cite{gillespie2018custodians, gruzd2023trust}, content creators \cite{hiaeshutter2021platform}, and viewers \cite{he2025survey}. We also recognized video-sharing platforms as important sites for ADHD communities to validate individual experiences and support each other, echoing prior work \cite{eagle2023, Leveille2024}.

Beyond insights from prior work, our research highlighted the unique video quality control challenges encountered by users with ADHD on VSPs. Compared to conventional social media which primarily uses text and images to deliver information \cite{zhao2019image}, videos provide richer information and a higher-level of stimulation \cite{Wittenberg2021}. Such a unique characteristic of video could make viewers with ADHD more inclined to seek interesting and stimulating content: Nine comments on TikTok specifically mentioned they were going to \textit{``scroll''} as they got bored, consistent with the ADHD tendency of sensation and stimulation seeking \cite{benedetto2019problematic}. Such a characteristic of ADHD could reduce the effectiveness of common misinformation management techniques such as debunking \cite{yousuf2021media}, if the presentation of such content was not well-designed. Furthermore, our findings also revealed that content posted by health professionals---who play an important role of debunking health-related misinformation on VSPs \cite{sharevski2024debunk}---were usually long and inaccessible to viewers with ADHD. Such a gap highlighted the importance of tailoring information quality-related content to the needs of viewers with ADHD on VSPs, such as converting misinformation debunking videos to short-form videos that were more ADHD-friendly. In addition, our findings revealed that creators’ video quality control practices were spread across various sections of the platform interface, including videos, descriptions, comments, and creator profiles. However, as comments indicated, viewers with ADHD could often be distracted by the platform’s complex visual layout and high information density (e.g., recommended videos). These distractions could make it difficult for viewers to notice important content quality control messages from creators, such as those included in video descriptions or creator profiles. This challenge highlighted the need for VSPs to design ADHD-friendly interfaces that present and prioritize key video quality control information in a more accessible and visible manner.

Our research also highlighted the unique needs of viewers with ADHD in communicating video quality issues. Unlike health content focusing on illnesses or disorders with clear physical symptoms requiring treatment or correction \cite{yoon2022understanding, li2022youtube}, activists in the ADHD community---and the broader neurodiverse community to which it belongs---have been fighting to validate the neurodiverse experiences and celebrate all individuals regardless of their neurological differences \cite{graby2015neurodiversity}. As a result, ADHD communities tend to embrace a more diverse set of identities regardless of their diagnostic status \cite{stenner2019adult, kazda2024attention}.  Given the complexity of ADHD community membership, our findings identified specific challenges encountered by some ADHD community members in video quality control practices. For example, as video creators reminded viewers to seek professional help and refrain from self-diagnosis, the self-diagnosed ADHD community members (many of whom highlighted their clinical diagnostic challenges in comments) inevitably had their identities questioned and stigmatized in the process. The dilemma between establishing a standard for video quality control and breaking the existing standard to truly embrace the spirit of neurodiversity movement makes video quality control in the ADHD community a particularly challenging task. Echoing calls from prior work \cite{eagle2023, Spiel2022}, we encourage future researchers to privilege community culture and member identities when investigating ways to carefully assess and indicate content quality on VSPs for communities like ADHD.

\subsection{Towards Safer and ADHD-inclusive VSPs}
\label{sub-sub-sec:design-accessibility}

Based on our findings, we derived \textbf{\textit{design considerations}} for creating safer, more trustworthy, and inclusive video platforms for individuals with ADHD.

\textbf{\textit{Facilitate Video Customizations and Simplification for ADHD.}} Our findings highlighted the challenges that ADHD viewers face during video watching, including having difficulties reducing visual/audio distractions and getting distracted by accelerated visuals when speeding up videos. This finding highlighted the need for future research to investigate video customization techniques for ADHD viewers. For example, computer vision technologies could be used to recognize and segment the visual information in videos (e.g., using object detection models \cite{khanam2024yolov11} to recognize the speaker and the presentation screens), and simplify a video with distracting visual components (e.g., removing distracting anime overlays with video inpainting techniques \cite{chang2019vornet}). Emerging video summarization models \cite{Alaa2024} also have the potential to shorten long videos into more concise and fast-paced video clips based on ADHD viewers' needs, as an alternative solution to speeding up videos. As over-simplification might lead to inconsistent information gain across viewers, we also encourage future research to consider how to mitigate such inconsistencies. For example, VLM models \cite{liu2024visual} could provide a summary of the hidden video content for viewers to review at the end.

\textbf{\textit{Leverage Community-driven Quality Control Efforts.}} Our findings revealed the collective efforts across VSPs, creators, and viewers in quality (and accessibility) control and improvement for ADHD-relevant videos. For example, many ADHD viewers leveraged comments to share alternative experiences that supplemented or challenged video content and added personal video breakdowns to make video presentations more accessible. %However, such efforts might not stand out easily in the abundance of comments. This finding highlights the needs of more effective leverage of the collective efforts, especially the comments section. 
Future research should consider how to better leverage such collective efforts to monitor and enhance video content quality and accessibility. For example, VSPs could provide a \textit{collective quality review} feature, using LLM models \cite{ji2023} to recognize and summarize comments and label video quality (e.g., flag a video if a large number of comments challenged the video content), or supplement the video content by adding different opinions from comments. Interaction techniques on how to best present this supplementary information without distracting ADHD viewers would also be a critical open question. More importantly, our findings indicated that comments could also include misleading or harmful information. %researchers should consider how to effectively filter our the misinformation in comments and summarize 
%We propose utilizing AI models to summarize \cite{Glickman2024} quality control-related comments and displaying the summary in a visible way (e.g., next to the video screen). Moreover, our findings also recognized the potential misinformation and harm that comments could bring, echoing prior work \cite{Anspach2018}. 
To alleviate this issue, AI summarization technology should also consider how to better distinguish and combat misinformation, for example, by gauging diverse perspectives (e.g., examining sub-level comments that deepen the discussion) and comprehensively presenting different opinions. %and remove inappropriate comments from the summary when necessary. 

\textbf{\textit{Reduce ADHD Efforts in Creative Work.}} Our findings identified the large proportion of creators with ADHD on VSPs sharing ADHD-relevant content and uncovered their challenges in taking extra steps to make their videos more credible and accessible (e.g., adding video chapters). More research is needed to enable a more streamlined and accessible content creation process for creators with ADHD. %neurodiverse creators' creative practices.
%\textbf{\textit{Content Quality Control.}} 
Current AI technology has the potential to automate many quality control efforts for creators. For example, video understanding models \cite{Bertasius2021} combined with LLMs could recognize the video content and generate automatic disclaimers (e.g., \textit{``This video is my personal experience only'')}, automatically extract and display creators' identity information (e.g., health professional in a specific field) from the video for higher visibility if they didn't disclose it explicitly in their profiles, or add video chapters and summaries automatically based on creator-specified prompts. Despite the potential, we recognize the risk of AI hallucination \cite{ji2023} and encourage future researchers to incorporate suitable techniques to reduce the impact caused by this issue, such as adding reminders of AI risks or enabling easy human verification.

\subsection{Limitations \& Future Work} Our research has limitations. Firstly, our work analyzed the experiences and perceptions of viewers with ADHD on VSPs via comment analysis, a method commonly used by prior literature for community understanding \cite{niu2022, hall2022did}. Nonetheless, we recognized the risks associated with this method, such as difficulty verifying the ADHD identities of the commenters. Despite the risks, we found value in analyzing the comment data, which comes from an anonymous and non-study setting. Such anonymity could afford a stronger inclination for self-disclosure \cite{ma2016anonymity} and more involved expressions \cite{de2014mental}, which are critical for uncovering the tension between users of different backgrounds and perspectives in ADHD communities (e.g., how viewers with ADHD questioned the authority of the creator). Leveraging the large-scale comment data, our work was also able to identify patterns of common challenges experienced by viewers of ADHD-relevant videos, and provide high-level pointers for future researchers to explore in depth (e.g., how to better support active ADHD help-seeking via VSPs). In future work, more empirical studies are needed to engage content creators and ADHD viewers directly to collect richer data that dives deep into users' personal experiences and choices. Moreover, our work lacked health expert inputs when discussing quality issues of health information. Future work could gauge perspectives from both the medical and ADHD communities for a more holistic analysis.
\section{Conclusion}
Our work contributed the first in-depth characterization and analysis of ADHD-relevant content on different VSPs. By carefully analyzing 373 videos and the corresponding comments, we characterized the unique creator demographics, content covered and presentation forms on each platform, identified the collective content quality control efforts and tensions across different user groups, and highlighted video accessibility issues for viewers with ADHD on each platform. Building on these insights, we proposed design guidelines on how VSPs could improve ADHD-relevant video sharing experiences to create a safer and more inclusive online environment for the ADHD community.

\begin{acks}
This work was partially supported by the University of Wisconsin—Madison Office of the Vice Chancellor for Research and Graduate Education with funding from the Wisconsin Alumni Research Foundation. 
\end{acks}

\bibliographystyle{ACM-Reference-Format}
\bibliography{main}


\end{document}